\documentclass[preprint,secnumarabic,a4paper,showkeys,aps,prb,endfloats*]{revtex4}
\usepackage{graphicx}

\newcommand{\etal}{\textit{et al.\hspace{0.5pc}}}

\begin{document}

\title[Nanofriction on nsc.]{Nanofriction behavior of cluster-assembled carbon films}
 \date{\today}
\author{A. Podest{\`a}, G. Fantoni, P. Milani}
\email{pmilani@mi.infn.it} \homepage{http://webcesid1.fisica.unimi.it/~labmilani/}
\affiliation{INFM - Dipartimento di Fisica, Universit{\`a} di Milano,\\
Via Celoria 16, 20133 Milano, Italy}
\author{M. Ragazzi, D. Donadio}
\affiliation{INFM - Dipartimento di Scienza dei Materiali, Universit{\`a} di Milano-Bicocca,\\
Via Cozzi 53, 20125 Milano, Italy}
\author{L. Colombo}
\affiliation{INFM - Dipartimento di Fisica, Universit\`a di Cagliari,\\ Cittadella
Universitaria 09042 Monserrato (CA), Italy}

\begin{abstract}
We have characterized the frictional properties of nanostructured (ns) carbon films
grown by Supersonic Cluster Beam Deposition (SCBD) via an Atomic Force-Friction
Force Microscope (AFM-FFM). The experimental data are discussed on the basis of a
modified Amonton's law for friction, stating a linear dependence of friction on load
plus an adhesive offset accounting for a finite friction force in the limit of null
total applied load. Molecular Dynamics simulations of the interaction of the AFM tip
with the nanostructured carbon confirm the validity of the friction model used for
this system. Experimental results show that the friction coefficient is not
influenced by the nanostructure of the films nor by the relative humidity. On the
other hand the adhesion coefficient depends on these parameters.
\end{abstract}


 \maketitle
\section{INTRODUCTION}

The understanding and control of friction, adhesion, lubrication and wear in
nanostructured systems is an essential requisite to validate the use of
nanomaterials for highly demanding structural applications \cite{bhu95,carp97}. The
rapidly growing number of applications based on microelectromechanical systems
(MEMS) and the new perspectives opened by the production of nano electromechanical
systems (NEMS) make necessary the development of an entirely novel class of
protective and lubricant coatings with improved mechanical properties at the
nanoscale \cite{mrsbull01}. Carbon-based materials have emerged as a promising class
of materials \cite{sull01}. Diamond and the various types of amorphous carbon show
interesting tribological properties such as high elastic moduli, good lubrication
properties, low stiction surfaces, etc. \cite{sull01}. These properties are
controlled by a combination of physico-chemical properties of the surfaces such as
sp2/sp3 ratio, surface roughness and porosity and by the presence of contaminant
layers \cite{sun99}.

Cluster-assembled carbon represents a novel nanostructured material obtained by the
deposition of clusters produced in a supersonic expansion \cite{mil01b}.
Cluster-assembled carbon films can be obtained starting from cluster mass
distributions that contain clusters with fullerene-like structure and/or linear and
planar structures \cite{don99}. Low-energy deposition (fractions of eV per atom)
drastically reduces fragmentation on the substrate allowing the films to be
structured at the nano- and mesoscale by keeping memory of the original cluster
distribution \cite{bar99b}.The mechanical properties of cluster-assembled carbon
films have been studied by Brillouin light scattering showing that these systems
have elastic properties similar to graphite (shear modulus) and Young modulus and
Poisson's ratio typical of a soft very porous material (note that the Poisson's
ratio is very close to zero or, in certain cases, even negative) \cite{cas01}.

The Atomic Force-Friction Force Microscope (AFM-FFM) has emerged as a powerful tool
for the characterization of the tribological properties of materials from the
micrometer down to the atomic scale \cite{carp97,ded00}. However, the use of a
nanometer-sized probe in nano-friction experiments carried out on corrugated samples
in humid environments causes the experimental conditions to be usually different
from both those encountered in typical ultra high-vacuum experiments, where flat
crystalline surfaces are investigated in a humidity and contaminants-free
environment, leading to a single-asperity contact, and those of macroscopic
tribology, where the contact regime is always multi-asperity like \cite{ada00}. A
number of parameters such as adhesion, surface and tip micro and nano-roughness,
load range, as well as tip radius and shape, influence the contact-friction regime.
Moreover, local corrugation on a scale larger than that of the probe can affect the
friction measurement \cite{koi97,sun00}. The case of nanostructured materials is
somehow peculiar since the typical size of the probe is close to or even comparable
with cluster size. In this case peculiar tip-sample interactions and topographic
effects should be expected.

In this paper we present the results of an AFM-FFM characterization of the
frictional properties of cluster-assembled carbon films. We have studied the
dependence of frictional parameters on both the relative humidity and the structural
composition of films deposited with different cluster mass distributions.

The experimental data are discussed on the basis of a modified Amonton's law for
friction \cite{bow50}, stating a linear dependence of friction on load plus an
adhesive offset accounting for a finite friction force in the limit of null total
applied load. A new procedure for the correction of the lateral force maps from the
contributions of the local tilt of the surface (the topographic correction) was
applied in order to extract intrinsic values of the frictional parameters
independent on surface roughness on a scale larger than that of the tip-sample
contact. In order to validate the use of the Amonton's law for the interpretation of
the experimental data, we have simulated the AFM tip-nanostructured carbon
interaction via molecular dynamics (MD) simulations. MD results of the nanofriction
experiment support the validity of the friction model used for this system.

\section{EXPERIMENTAL DETAILS}

\subsection{Deposition of nanostructured carbon}

The use of supersonic beams of clusters for deposition of thin films has attracted a
large interest from more than two decades\cite{mil99}. This technique consists in
preparing clusters in the gas phase diluted in a carrier light inert gas, typically
Helium, and letting the mixture expand through a nozzle in high vacuum so that a
very collimated, intense supersonic beam is produced. The cluster beam is
intercepted by a suitable substrate in order to deposit a thin film. In the case of
carbon, the kinetic energy per atom in the clusters (below 0.4 eV/atom) is smaller
than the binding energy per atom avoiding massive fragmentation: the resulting film
has a structure at the nanoscale keeping the memory of the nanometer-sized building
blocks used for the assembling.

Nanostructured carbon films have been deposited from a supersonic cluster beam
produced by a pulsed microplasma cluster source (PMCS) as described in detail in
Ref. \onlinecite{bar99}. With normal PMCS operation conditions, the cluster beam is
characterized by a log-normal cluster mass distribution peaked at about 500
atoms/cluster and extending to several thousands atoms per cluster. We have
controlled and varied the cluster mass distribution and deposition rates by
exploiting aerodynamic focusing effects \cite{pis01}. Using the standard cluster
mass distribution described above, we have deposited films with thicknesses of
several hundreds of nanometers on silicon substrates. The deposition rate was 4-5
nm/min and the density of the films was 0.8-0.9 g/cm$^3$. With a suitable nozzle
configuration we have produced a beam depleted from clusters with diameters roughly
larger than 2 nm \cite{pis01}. These conditions produce films with densities of
1.2-1.3 g/cm$^3$ at a deposition rate of 5 nm/sec.

Raman analysis have shown that in films grown with large clusters the graphitic
$sp2$ bonding, due to the large number of cage-like particles, is more pronounced
than in films grown with small clusters, whose structure resembles more that of the
amorphous carbon. In the following, we shall refer to films grown with the two
different nozzles as to the films grown with small and large clusters, accordingly.

\subsection{AFM-FFM}

The atomic force microscope is a Nanoscope Multimode IIIa from Digital Instruments
with phase extender and Signal Acquisition Module (SAM). We scanned different points
of the samples, with scan size 500 nm, sliding velocity typically 1 $\mu m/s$. We
used rectangular cantilevers 450 $\mu m$ long with silicon tip with radius 5-40 nm.
The AFM can be housed in a sealed chamber connected to a humidifier to work in
controlled humidity and atmosphere (typically in dry and wet nitrogen). A second PC
is used to remotely control the external applied load in friction measurements. The
application of the external load is synchronized via the reading of the end-line and
end-frame triggers from the microscope. We can acquire a complete friction vs. load
curve in a single AFM scan of typically 512 lines, 512 points per line. This allows
performing many measurements in each session, fast and reliably. The experimental
data are extracted in ASCII format and processed via dedicated software.

\section{\label{sec:MD_setup}MOLECULAR DYNAMICS SIMULATIONS}

The theoretical investigation has been carried out by classical MD simulations based
on the Tersoff potential \cite{tersoff}. This computational framework has proved to
be reliable for the investigation of the structural properties of carbon based
materials \cite{terscarb} and particularly of nanostructured cluster assembled
carbon films. The growth of nanostructured carbon films by supersonic cluster beam
deposition on a (001) diamond substrate has been simulated by a protocol described
in Ref. \onlinecite{don99} which accounts well for the experimental deposition
process. The simulation cell is a slab with periodic boundary conditions applied in
the plane orthogonal to the growth direction, the substrate is four layers thick and
its bottom layer is rigid, while the second and third layers are thermostat. The
dynamics of the atoms of the forth layer is newtonian as that of the impinging
clusters. Two samples have been produced with the same growth conditions (cluster
kinetic energy, vibrational temperature, substrate temperature) but different size
distribution of the precursors: sample (A) has been obtained from a cluster beam
mostly containing small precursors (1 to 23 atoms per cluster) and sample (B) has
been grown mainly from larger precursors (46 to 120 atoms per cluster). The
structural properties of the two sample, which are strongly influenced by the
features of the precursors, are discussed elsewhere \cite{don99}.

\begin{figure}[t]
\centering
\includegraphics[scale=0.8]{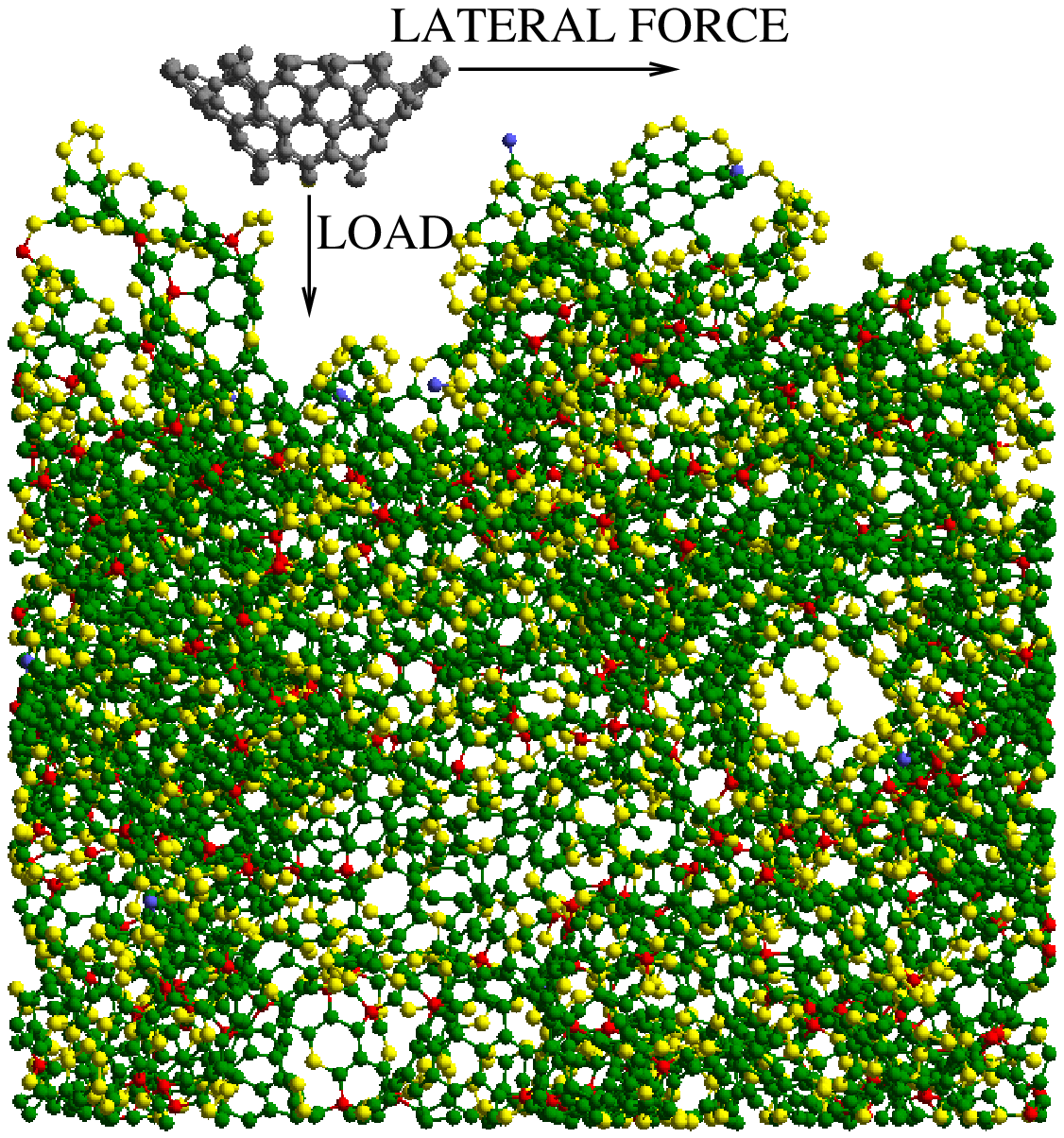}
\caption{\label{th1}}
\end{figure}

Fig. \ref{th1} illustrates the molecular dynamics model for the FFM experiment: a
small crystalline diamond tip slides on the surface of the nanostructured carbon.
The tip consists of 178 carbon atoms arranged in a truncated pyramid shape, obtained
by cutting an ideal diamond structure crystal along (111) and (001) directions and
letting it relax at room temperature. Tip radius is about 8 \AA. Although silicon
tips are used in experimental measurements we have chosen to perform nanofriction
simulations with a stiffer carbon tip, in order to avoid the complete wearing of the
tip even at relatively low normal loads, due to its small size. Thus it has been
possible to perform  FFM simulations with loads up to 30 nN.
Since the thickness of the samples exploited for nano-friction simulations is about
60 \AA , in order to reduce computational costs, a 20 \AA\ thick layer of the film
is fixed, the atoms belonging to the 2 \AA\ thick layer above it are thermostat and
the dynamics of the other atoms of the film is Newtonian. The two top layers of the
tip are rigid and they are subject to the external forces that mimic the action of
the AFM cantilever. The overall external force acting on the tip has been decomposed
into three independent components: (i) an elastic dragging force that produces the
sliding motion of the tip, (ii) a constant load perpendicular to the sliding plane
and (iii) an elastic torsion force that keeps the tip vertical with respect to the
sliding plane. The frictional force is calculated at every timestep as the reaction
to the elastic dragging force and it is averaged over six paths $20$ \AA\ long.

Since the formation of chemical bonds between the atoms of the AFM tip and those of
the film is quite unlikely in such AFM-FFM experiments, the film--tip interaction is
ruled by Van der Waals forces. Therefore the film--tip interaction is modelled by a
two-body modified Morse potential given by:
\begin{equation}
  V(r_{ij})=D(e^{-2\alpha(r_{ij}-d-r_{0})}-2e^{\alpha(r_{ij}-r_{0})}),
\end{equation}
where $r_{ij}$ is the distance between two atoms. As for $D$ and $\alpha$ parameters
we have chosen the values fitted by Cheong et al. \cite{cheong}: namely $D=0.435$ eV
and $\alpha=46.487$ . The parameter $d$ has been introduced in order to increase the
equilibrium distance and to reduce the binding energy between the tip and the film.
This reproduces better the actual experimental conditions: as the FFM measurements
are usually performed in air, we have to assume that the surface-tip interaction is
modified by hydrogen passivation of the dangling bonds and by the presence of
surface lubricants or contaminants. Hence, in principle, by tuning this parameter
one may study the effects of lubricants and contaminants without explicitly
introduce them. We have chosen $d=0.011$ nm, by performing sliding friction
simulations on the (001) non reconstructed surface of diamond and comparing the
calculated lateral force to those measured in Ref. \onlinecite{mate}.

\section{Results and discussion}

\subsection{AFM-FFM measurements}

Our characterization protocol assumes a modified Amonton's law for friction, i.e. a
linear dependence on total load plus an offset representing the zero-load friction
force in analogy with the single-asperity JKR contact model\cite{joh71}:
\begin{equation}
f=\mu N+c \label{eq:mod_amonton}
\end{equation}
Here $N$ represents the \textit{total} applied load in the direction perpendicular
to the surface, including the contribution of adhesion $A$, and $c$ accounts for a
zero-load friction force.

We apply a procedure, described in details elsewhere \cite{pod02a,pod02b}, to
correct lateral force maps from the spurious contributions due to the presence of a
local tilt of the surface. Actually, in the case of a locally tilted surface, the
measured forces in the directions parallel and perpendicular to the AFM reference
plane do not necessarily coincide with the forces acting parallel and
perpendicularly to the sample surface, which actually define the friction
coefficient and the friction vs. load characteristics of the interface under
investigation. The topographic correction is necessary in order to extract
quantitative and accurate information from corrugated sample and to compare results
from different samples.

We have investigated frictional properties of films grown with large and small
clusters in dry and humid nitrogen environment.
\begin{figure}[t]
\centering
\includegraphics[scale=0.9]{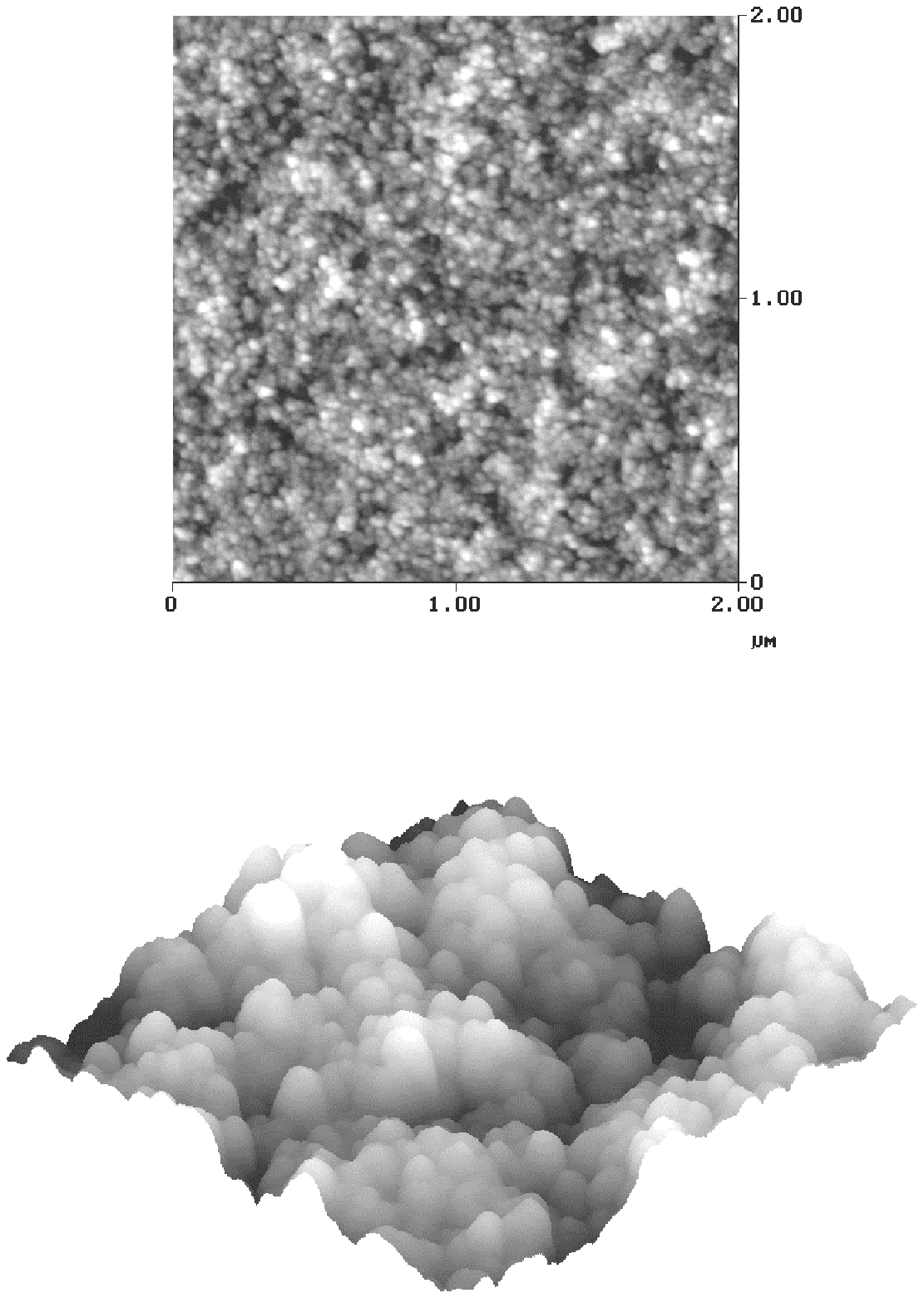}
 \caption{\label{fig:nsc_art}}
\end{figure}
In Fig. \ref{fig:nsc_art} are shown two AFM pictures at different magnifications of
a 250 nm thick ns-carbon film. Typical film morphology consists in a fine raster of
grains with typical diameter of 10-20 nm. Films grown with large clusters are
usually rougher than films grown with small clusters with the same thickness.
However, our protocol automatically corrects the friction maps for the topographic
contributions, allowing the comparison of the results obtained on different samples.

\begin{figure}[t]
\centering
\includegraphics[scale=0.8]{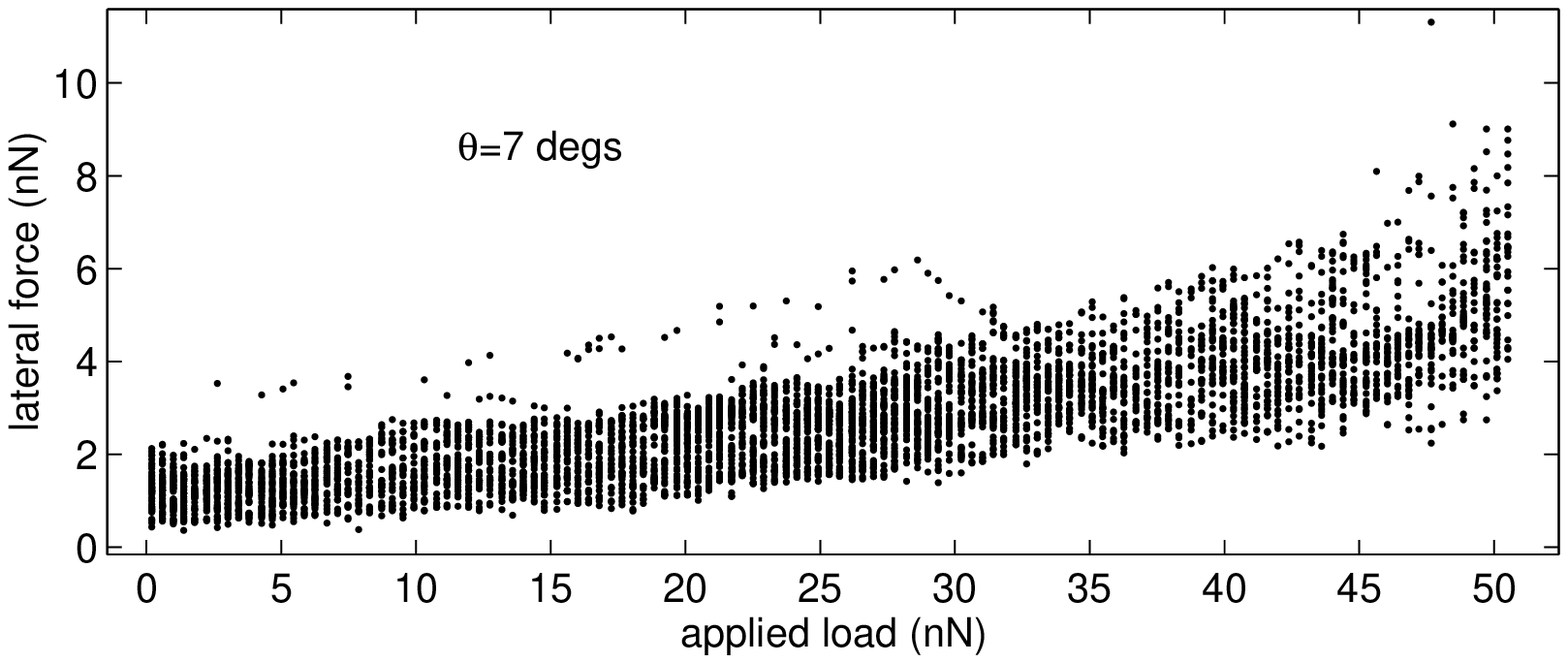}
 \caption{\label{fig:lateral_nsc}}
\end{figure}

In Fig. \ref{fig:lateral_nsc} is shown a typical lateral force vs. applied load
curve measured on ns-carbon. This curve is obtained extracting from the original
lateral force-load dispersion the subset of all the lateral force-load pairs
corresponding to the same slope in the topographic map. Each of such curves is
processed separately in order to apply the topographic correction.

\begin{figure}[t]
\centering
\includegraphics[scale=0.9]{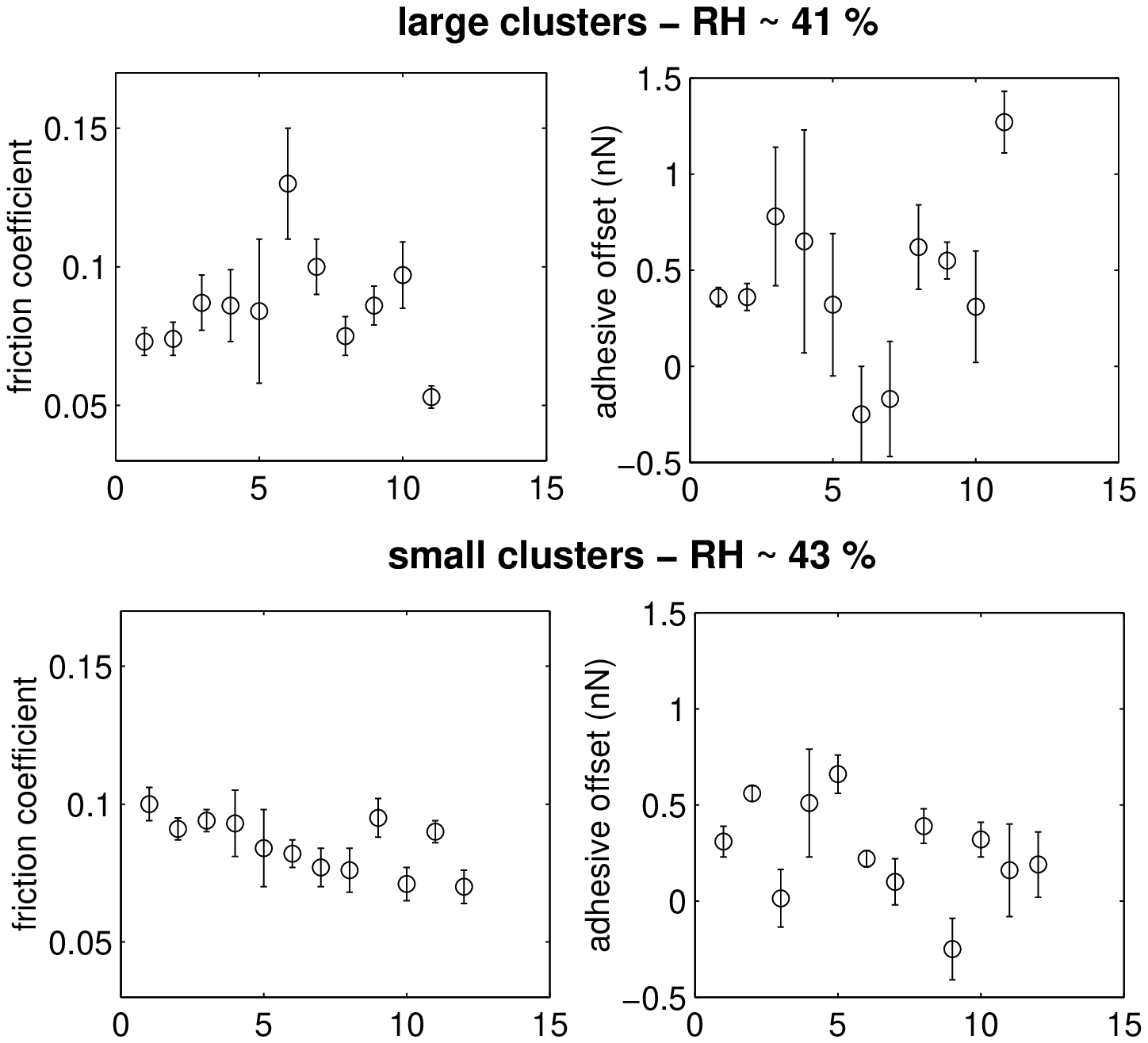}
\caption{\label{fig:nsc_friction_humid}}
\end{figure}

We show in Fig. \ref{fig:nsc_friction_humid} the measured \textit{dynamic} friction
coefficients and friction offsets for the film grown with large clusters and grown
with small clusters accordingly, measured in ambient condition
(RH$\sim40\%$)\footnote{One should keep in mind that we have measured the friction
coefficient and adhesive offset of the pair (ns-carbon, Silicon AFM tip). Using AFM
tips with different composition can lead in principle to different friction
coefficients and adhesive offsets.}.
\begin{figure}[t]
\centering
\includegraphics[scale=0.9]{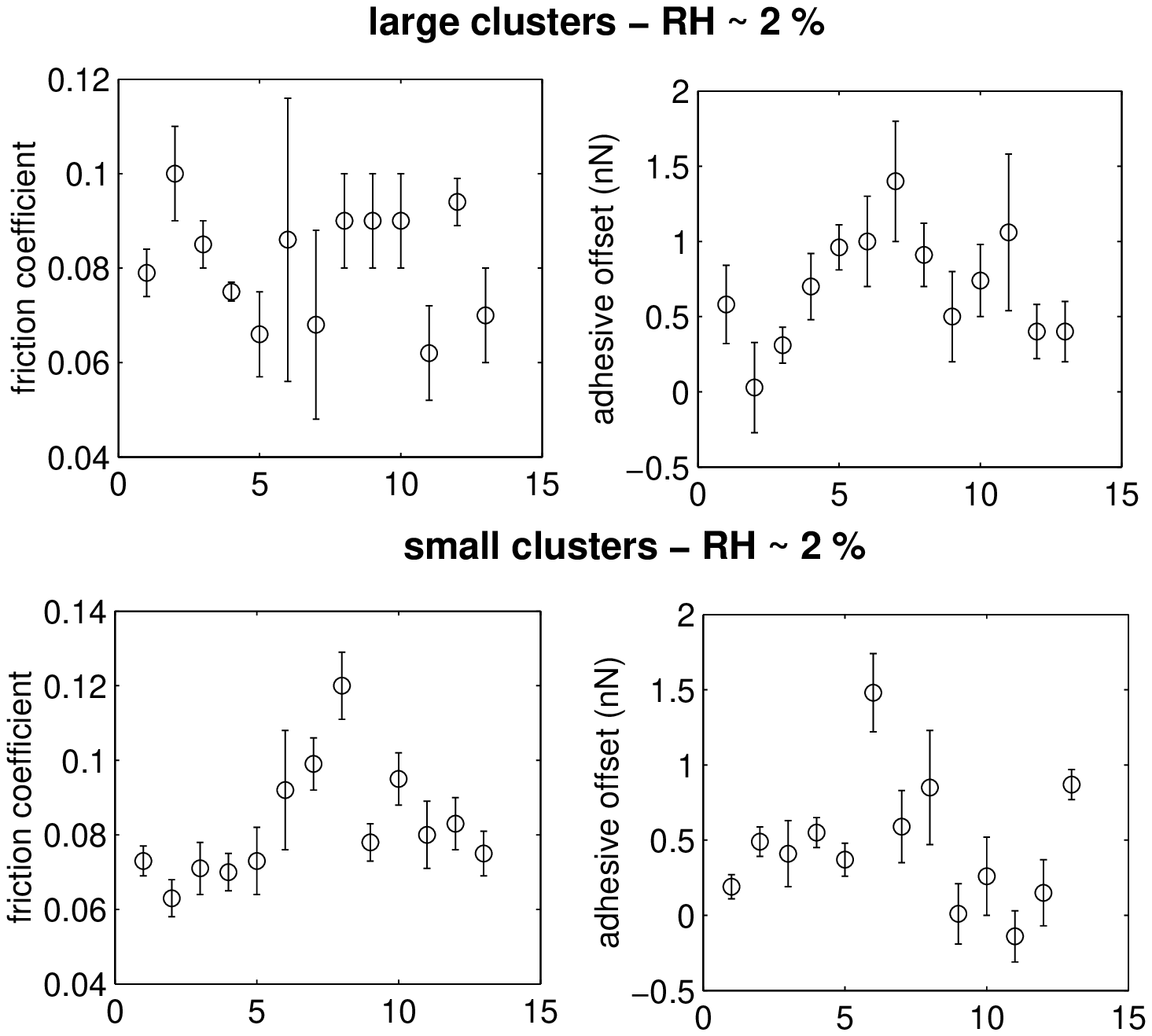}
 \caption{\label{fig:nsc_friction_dry}}
\end{figure}
The average values for both $\mu$ and $c$ are:
\begin{eqnarray}
  \mu^{humid}_{large} = 0.086\pm0.013, & \hspace{.2cm} c^{humid}_{large} =
  0.44\pm0.29 \, nN  \nonumber \\
  \mu^{humid}_{small} = 0.085\pm0.007, & \hspace{.2cm} c^{humid}_{small} =
  0.26\pm0.15 \, nN \label{eq:muc_humid}
\end{eqnarray}
We also studied the frictional behavior of ns-carbon film in dry environment
(RH$\sim 2\%$). The results are shown in Fig. \ref{fig:nsc_friction_dry}. The
averaged values are:
\begin{eqnarray}
  \mu^{dry}_{large} = 0.081\pm0.012,  & \hspace{.2cm} c^{dry}_{large} =
  0.69\pm0.28\, nN \nonumber \\
  \mu^{dry}_{small} = 0.082\pm0.008, & \hspace{.2cm} c^{dry}_{small} =
  0.47\pm0.20\, nN \label{eq:muc_dry}
\end{eqnarray}
The errors shown in Eqs. \ref{eq:muc_humid} and \ref{eq:muc_dry} are calculated as:
$\sigma=1/N\sqrt{\sum{\sigma^2_i}}$, where $\sigma_i$ are the errors associated to
each measurements.

\subsection{MD results}

\begin{figure}[t]
\centering
\includegraphics[scale=0.35]{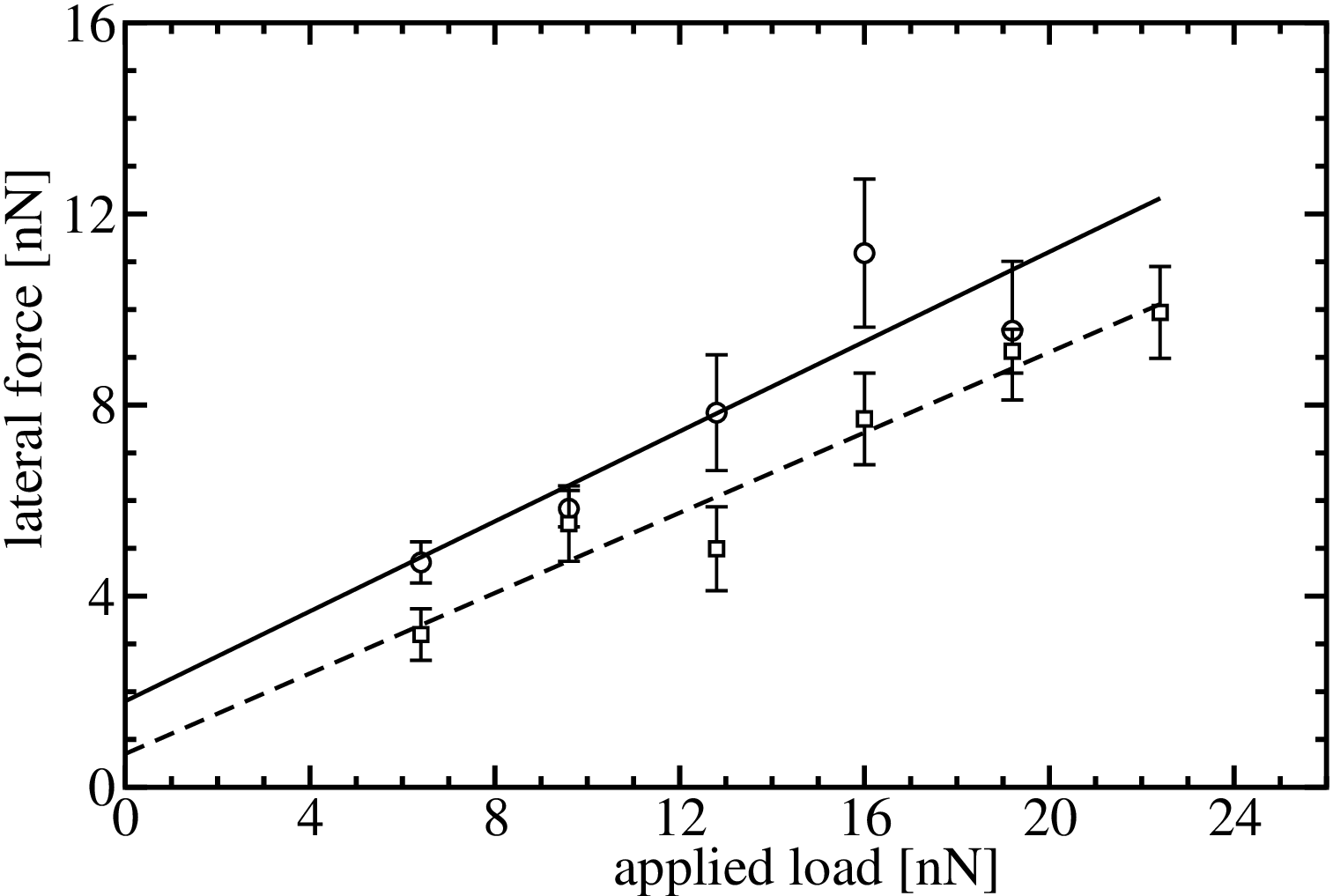}
 \caption{\label{th2}}
\end{figure}

Exploiting the model setup described in Section \ref{sec:MD_setup}, we have
performed several nanofriction simulations, calculating the values of lateral
friction force as a function of the applied normal load, both for sample (A) and for
sample (B). Each simulated measurement at a fixed load, consists of six independent
scannings 25 \AA\ long. The friction force for each scansion is obtained by the
average of the lateral force calculated at every time-step of the simulation and the
values reported in Fig. \ref{th2} are calculated by averaging over the six
scannings. This procedure provides a sufficient amount of data so to devise a linear
trend in the lateral force as a function of applied load and to calculate the slope,
i.e. the friction coefficient, and the offset of the curves according to the
modified Amonton's law discussed above.
The friction coefficient and offset values thus obtained are:
\begin{eqnarray}
  \mu^{A} = 0.47\pm 0.110,  & \hspace{.2cm} c^{A} = 1.81\pm\ 1.43 , nN \nonumber \\
  \mu^{B} = 0.42\pm 0.053, & \hspace{.2cm} c^{B} = 0.70\pm\ 0.59, nN \label{eq:mu_th}
\end{eqnarray}
Simulation have been performed with loads up to 22.6 nN for sample (A) and 19.2 nN
for sample (B). Beyond these threshold values, combined wearing of the tip and of
the film occurs and lateral force rapidly increases above the linear behavior
observed so far.

No topographic correction to the theoretically calculated friction vs. load curves
has been applied, since the method used to calculate each single point of the curves
averages over the scanned profile and this should already account for the
topographic effects. Moreover, the average tilt angle of the MD profiles is in
general smaller than 15 degrees, making the topographic correction negligible.

\subsection{Discussion}

The observed lateral force vs. applied load curves for ns-carbon are linear, except
for very low loads, close to the pull-off limit, and for loads larger than about 30
nN (Fig. \ref{fig:lateral_nsc}). Moreover, in the limit of the experimental error,
the measured values of the friction offset are definitely different from zero. These
observations confirm that the friction law of a silicon tip on cluster-assembled
carbon is well represented by an Amonton's like equation, and in addition that a
zero-load offset must be included in the friction model.

MD results further support our model based on a modified Amonton's law. Simulations
show a linear trend of the friction vs. load curve with a finite offset, different
from zero in the limit of the experimental error. The quantitative discrepancy
between theoretical and experimental results is not surprising and it can be
attributed to the parameterization of the interaction potentials and in the choice
of using carbon tips in the simulations of the FFM experiments.

For loads larger than about 30 nN we have evidence of the onset of non-linear trends
in friction vs. load curves, probably related to wear and indentation during
scanning. The phenomenon of wearing of the ns-carbon film has been directly observed
in MD simulations at loads larger than 30 nN.

It has been reported that the values of the friction coefficients measured in
successive scans are likely to be different \cite{carp97,koi96}. This is currently
attributed to modification of the tip-sample interface after repeated sliding due to
material removal, tip wear and blunting and even structural modifications. Blunting
of the tip is a very unlike event in this case because Silicon is much harder than
cluster-assembled carbon. The observed fluctuations in the measured parameters (see
Figs. \ref{fig:nsc_friction_humid} and \ref{fig:nsc_friction_dry}) could be
attributed to dynamic modifications of the contact area during sliding because of
the contamination of the AFM tip, which may pick up carbon during scanning,
particularly at higher loads, and even indent the film, which is much softer than
the tip.

The strongest fluctuations are observed for the experimental friction offset $c$.
This is not surprising, because this parameter is expected to be highly dependent
not only upon environmental conditions (such as relative humidity and temperature),
but also upon tip shape and contamination, these being the parameters directly
influencing adhesion. This fact is clearly represented by the large errors in the
measured adhesive offset. While the relative error for the friction coefficient are
below 15$\%$, which is basically the limit set by the accuracy of the calibration
procedure for the force constants of the cantilever\cite{gibson96}\footnote{The
statistical error in each measurement is in general very small, because with our
procedure we may acquire up to 512x512 friction-load pairs in each image and many
images fast and reliably.}, those of the adhesive offset are as large as 60$\%$. In
the absence of an accurate characterization of tip shape and size however it is not
possible to definitely individuate the causes of the observed changes in the
friction offset $c$. The investigation of adhesive properties of cluster-assembled
carbon should be more reliable through the analysis of force vs. distance curves, or
even via the study of thermal fluctuations of the cantilever close to the
nanostructured surface, provided a reliable model of the tip-sample interaction
and/or contact.

As confirmed also by MD simulations, in the limit of the experimental error, the
nanostructure of the films seems to have very weak or no influence on the friction
coefficient. In analogy to the macroscale, one could expect that films grown with
large clusters and hence containing a larger number of well-organized graphitic
regions should show a lower friction coefficient \cite{grill97}. This is not
observed.

The friction coefficient is also found to be independent on relative humidity. Null
or weak dependence of friction coefficient and adhesion on the relative humidity was
reported in literature for carbon-based materials\cite{bing94,koi96,bhu00}. One
possible explanation for the behavior of our films in dry environment is the
hydrophilic nature of cluster-assembled carbon \cite{ostro02}. The low density and
large porosity of these systems in conjunction with a non-negligible wettability may
favor the formation of a water layer lasting even in dry nitrogen with the formation
of a water meniscus between the tip and the surface

Friction coefficients measured on carbon-based materials span from 0.009 (for HOPG
in air) to 1 for DLC in UHV \cite{grill97}. A numerical comparison with
nanostructured carbon would not be of great interest because of the large scattering
of the results and the arbitrariness of the calibration procedures adopted. What is
in general observed is that all carbon-based materials are better lubricants in the
presence of surface contamination, while in very clean and dry environments the
covalent interaction between sample and probe (usually Silicon or Diamond) leads to
higher friction coefficients. In the case of cluster-assembled carbon we do not
observe an increase of the friction coefficient upon removal of the environmental
humidity.

The adhesive offset $c$ is found experimentally to decrease in humid environment.
The decrease of the adhesion force with increasing relative humidity was observed in
other systems \cite{bing94}. This is explained by a decrease of the capillary
attractive force exerted by the water meniscus in the tip-surface gap, with higher
relative humidity, upon a certain threshold, depending on the material. This could
explain also the larger adhesion coefficient observed in the films grown from large
clusters. This system with a more pronounced graphitic character may have a smaller
wettability compared to that grown by small clusters \cite{ostro02}.

In the simulations the effect of humidity is poorly reproduced by the introduction
of the parameter $d$ in the empirical potential describing the tip-film interaction
since it is impossible to establish a clear relation between the relative humidity
and the value of $d$. Anyway, we have observed that in completely dry conditions,
corresponding to $d=0$ \AA, strong adhesion of the tip on the surface occurs,
resulting in the complete wearing of the tip and in large deformation of the scanned
surface, even for loads as small as 5 nN.

Recently Buzio \etal reported a friction characterization on cluster-assembled
carbon films produced by our technique \cite{buz02}. They have found a non-linear
dependence of friction on load, well fitted by the Hertzian-plus-offset model
\cite{schw97}, using a silicon tip of $\sim 30$ nm and scan length of 50 nm. On the
other hand, a linear dependence of friction on load is found by scanning on a larger
area (1x1 $\mu$m$^2$) \cite{gne00}. The authors attributed the non-linear dependence
to a possible passivation of the sample surface and to the presence of a water
layer, which would induce a transition from a multi-asperity to a single-asperity
contact (i.e. from a linear to a power-law dependence) as explained in the framework
of the composite-tip model \cite{put95}. For larger scan length and relatively
larger scan velocity, the tip-sample interface is broken and re-formed rapidly and
the smoothing action of water and impurities is less effective. This hypothesis is
confirmed by our measurements apart from differences observed in the numerical
values of the friction coefficients.

\section{CONCLUSIONS}

We have characterized by FFM-AFM the tribological properties of nanostructured
carbon films grown by supersonic cluster beam deposition. We have found that the
friction behavior at the nanoscale can be described by a modified Amonton's law
demonstrating that a nanometer-sized contact not necessarily leads to a
single-asperity friction regime. Molecular Dynamics simulations support our
interpretation and show a qualitative agreement with the experiments. Our results
show that the presence of water and the high reactivity of $sp2$-$sp3$ material
plays an important role in determining a multi-asperity like contact also at these
very small scales. Although numerical comparison with other carbon-based systems
must be considered with high caution, cluster-assembled carbon shows a low friction
coefficient even in a dry environment. The nanoscale structure of the films seem to
have a negligible influence on the friction at the observed scales. This suggests
that the tribological behavior of a nanostructured solid is "scale-sensitive" and
one should always consider the scale factor before making comparison or
extrapolations.

\begin{acknowledgments}
Partial financial support from Italian Space Agency is acknowledged.
\end{acknowledgments}

\newpage
\section*{Figure captions}
\begin{enumerate}
\item[Fig. \ref{th1}.]Representation of the model used for AFM-FFM molecular dynamics
                      simulations. Only a $\sim$50 \AA\ thick top layer of nanostructured carbon
                      film is shown. Color map reflects the coordination of atoms in the ns-carbon film:
                      Blue, 1-fold coordinated; yellow, 2; green, 3; red, 4.
\item[Fig. \ref{fig:nsc_art}.] AFM pictures of a 250 nm thick ns-carbon film at different magnifications.
                               The scan size is 2 $\mu$m (top) and 500 nm (bottom). The vertical color scale
                               is 100 nm. The higher magnification picture clearly shows the surface granularity
                               of such nanostructured films.
\item[Fig. \ref{fig:lateral_nsc}]Lateral force vs. (external) applied load curve measured on
                                 ns-carbon.
\item[Fig. \ref{fig:nsc_friction_humid}]Measured friction coefficients $\mu$ and friction offset $c$ for films
                                        grown with large clusters and grown with small clusters in humid nitrogen (RH$\sim 40\%$).
\item[Fig. \ref{fig:nsc_friction_dry}]Measured friction coefficients $\mu$ and friction offset $c$ for films grown with large clusters and
                                      grown with small clusters in dry nitrogen (RH$\sim 2\%$).
\item[Fig. \ref{th2}]Lateral force vs. applied load values and fitting from MD
           simulations, for sample (A) (circles and solid line) and for sample (B)
           (squares and dashed line). Error bars represent the standard deviation
           of the lateral force calculated for each scansion at a given load.
\end{enumerate}

\end{document}